\documentstyle[12pt]{article}

\font\twlgot =eufm10 scaled \magstep1
\font\egtgot =eufm8
\font\sevgot =eufm7
\font\twlmsb =msbm10 scaled \magstep1
\font\egtmsb =msbm8
\font\sevmsb =msbm7

\newfam\gotfam
\def\pgot{\fam\gotfam\twlgot}
\textfont\gotfam\twlgot
\scriptfont\gotfam\egtgot
\scriptscriptfont\gotfam\sevgot
\def\got{\protect\pgot}
\newfam\msbfam
\textfont\msbfam\twlmsb
\scriptfont\msbfam\egtmsb
\scriptscriptfont\msbfam\sevmsb
\def\Bbb{\protect\pBbb}
\def\pBbb{\relax\ifmmode\expandafter\Bb\else\typeout{You cann't use
Bbb in text mode}\fi}
\def\Bb #1{{\fam\msbfam\relax#1}}

\newcommand{\gS}{{\got S}}

\newcommand{\gG}{{\got G}}
\newcommand{\gC}{{\got C}}
\newcommand{\gM}{{\got M}}
\newcommand{\gd}{{\got d}}
\newcommand{\gR}{{\got R}}
\newcommand{\gA}{{\got A}}

\def\thebibliography#1{\section*{References}\list
   {[\arabic{enumi}]}{\settowidth\labelwidth{#1}\leftmargin\labelwidth
     \advance\leftmargin\labelsep
     \usecounter{enumi}}
     \def\newblock{\hskip .11em plus .33em minus .07em}
     \sloppy\clubpenalty4000\widowpenalty4000
     \sfcode`\.=1000\relax}

\def\op#1{\mathop{\fam0 #1}\limits}

\newcommand{\glos}[1]{\bigskip{\bf #1}\medskip}

\newcommand{\id}{{\rm Id\,}}
\newcommand{\Ker}{{\rm Ker\,}}

\newcommand{\beq}{\begin{equation}}
\newcommand{\eeq}{\end{equation}}
\newcommand{\ben}{\begin{eqnarray}}
\newcommand{\een}{\end{eqnarray}}
\newcommand{\be}{\begin{eqnarray*}}
\newcommand{\ee}{\end{eqnarray*}}
\newcommand{\bea}{\begin{eqalph}}
\newcommand{\eea}{\end{eqalph}}
\newcommand{\cA}{{\cal A}}

\newcommand{\cP}{{\cal P}}
\newcommand{\cD}{{\cal D}}
\newcommand{\cR}{{\cal R}}

\newcommand{\cM}{{\cal M}}
\newcommand{\cJ}{{\cal J}}
\newcommand{\cE}{{\cal E}}

\newcommand{\cO}{{\cal O}}
\newcommand{\cG}{{\got g}}
\newcommand{\cV}{{\cal V}}
\newcommand{\cB}{{\cal B}}
\newcommand{\cK}{{\cal K}}
\newcommand{\cQ}{{\cal Q}}
\newcommand{\cZ}{{\cal Z}}

\newcommand{\bt}{\beta}
\newcommand{\dl}{\delta}
\newcommand{\la}{\lambda}

\newcommand{\f}{\phi}

\newcommand{\Om}{\Omega}
\newcommand{\m}{\mu}
\newcommand{\n}{\nu}
\newcommand{\g}{\gamma}

\newcommand{\th}{\theta}

\newcommand{\di}{{\rm dim\,}}

\newcommand{\dif}{{\rm Diff\,}}

\newcommand{\si}{\sigma}
\newcommand{\Si}{\Sigma}
\newcommand{\w}{\wedge}

\newcommand{\wh}{\widehat}
\newcommand{\ol}{\overline}
\newcommand{\dr}{\partial}

\newcommand{\ot}{\otimes}
\newcommand{\ap}{\approx}
\newcommand{\ve}{\varepsilon}

\newcounter{eqalph}
\newcounter{equationa}
\newcounter{theorem}
\newcounter{remark}
\newcounter{proposition}
\newcounter{lemma}
\newcounter{corollary}
\newcounter{definition}

\newenvironment{eqalph}{\stepcounter{equation}
\setcounter{equationa}{\value{equation}}
\setcounter{equation}{0}

\begin{eqnarray}}{\end{eqnarray}\setcounter{equation}{\value{equationa}}}

\def\theremark{\arabic{remark}}

\def\thedefinition{\arabic{definition}}

\newcommand{\mar}[1]{}

\begin{document}
\hbox{}

{\parindent=0pt

{\large\bf What is geometry in quantum theory}
\bigskip

{\sc Gennadi Sardanashvily}

Department of Theoretical Physics,  Moscow State
University, 117234 Moscow, Russia

E-mail: sard@grav.phys.msu.su

Web: http://webcenter.ru/$\sim$sardan/ 
\bigskip

{\sc Giovanni Giachetta}

Department of Mathematics and Informatics, 
University of Camerino, 62032 Camerino (MC), Italy 

E-mail: giovanni.giachetta@unicam.it
\bigskip

\begin{small}
{\bf Abstract}.
In this scientific preface to the first issue of {\it
International Journal of Geometric Methods in Modern
Physics}\footnote{\bf
Web: http://www.worldscinet.com/ijgmmp/ijgmmp.shtml}, we
briefly survey some peculiarities of geometric
techniques in quantum models.

\end{small}

}

\bigskip
\bigskip

Contemporary quantum theory meets an explosion of
different types of quantization. Some of them
(geometric quantization, deformation quantization,
noncommutative geometry, topological field theory etc.) 
speak the language of geometry, algebraic and differential
topology. We do 
not pretend for any comprehensive analysis of these
quantization techniques, but aims to formulate and illustrate
their main peculiarities. 
As in any survey, a selection of topics has
to be done, and we apologize in advance if some relevant
works are omitted.

Geometry of classical mechanics and field theory is
mainly differential geometry of finite-dimensional smooth
manifolds, fiber bundles and Lie groups.  
The key point why geometry plays a prominent role in
classical field theory lies in the fact that it enables one to
deal with invariantly defined objects. Gauge theory has shown
clearly that this is a basic physical principle. At first,
a pseudo-Riemannian metric has been identified to a
gravitational field in the framework of Einstein's General
Relativity.  In 60-70th, one has observed that connections on a
principal bundle provide the
mathematical model of classical gauge potentials [1-3].
Furthermore, since the characteristic classes of principal
bundles are expressed in terms of the gauge strengths, one can
also describe the topological phenomena in classical gauge
models
\cite{eguchi}. Spontaneous symmetry breaking and Higgs
fields have been explained in terms of reduced $G$-structures
\cite{nik}. A gravitational field seen as a pseudo-Riemannian
metric exemplifies such a Higgs field \cite{iva}. In a general
setting, differential geometry of smooth fiber bundles gives
the adequate mathematical formulation of classical field
theory, where fields are represented by sections of fiber
bundles and their dynamics is phrased in terms of jet manifolds
\cite{book}. Autonomous classical mechanics speaks the geometric
language of symplectic and Poisson manifolds
[8-10]. Nonrelativistic time-dependent mechanics
can be formulated as a particular field theory on fiber bundles
over $\Bbb R$ \cite{book98}.

At the same time, the standard 
mathematical language of
quantum mechanics and perturbative field theory, except
gravitation theory, has been long far from geometry. In the
last twenty years, the incremental development of new  
physical ideas in quantum theory (including 
super- and 
BRST symmetries, geometric and deformation
quantization, topological field theory, anomalies,
noncommutativity, strings and branes) has called into play
advanced geometric techniques, based on the deep interplay
between algebra, geometry and topology.

In the framework of algebraic quantization,
one associates to a classical system a certain 
(e.g., von Neumann, $C^*$- or $Op^*$-) algebra 
whose different representations are studied. 
Quantization techniques under
discussion introduce
something new. Namely, they can provide
nonequivalent quantizations of a classical system
corresponding to different values  of some topological and
differential invariants. For instance,  a symplectic 
manifold
$X$ admits a set of nonequivalent star-products indexed by
elements of the cohomology group
$H^2(X)[[\la]]$ \cite{nest,gutt}. Thus, one
may associate to a classical system different underlying 
quantum models.  Of course, there is a
question whether this ambiguity is of physical or only
mathematical nature. From the mathematical viewpoint, one may
propose that any quantization should
be a functor between classical and quantum categories (e.g.,
some subcategory of Poisson manifolds on the classical side and
a subcategory of
$C^*$-algebras on the quantum side) \cite{lands02}.
From the physical point of view,  dequantization becomes
important.

There are several examples of {\it sui generis} dequantizations.
For instance,  Berezin's quantization
\cite{engl} in fact is dequantization. One can also think of
well-known Gelfand's map as being dequantization of a
commutative
$C^*$-algebra $\cA$ by the algebra of continuous complex
functions vanishing at infinity on the spectrum of $\cA$.
This dequantization has been generalized to
noncommutative unital
$C^*$-algebras
\cite{cir,kawa00}. The concept of the of the strict
$C^*$-algebraic deformation quantization implies an appropriate
dequantization when $\hbar\to 0$ \cite{rieff,lands282}. In
Connes' noncommutative geometry, dequantization of the spectral
triple in the case of a commutative algebra
$C^\infty(X)$ is performed in order to restart the original
differential geometry of a spin manifold
$X$ \cite{conn,ren2}. One can also treat the method of an 
effective action in perturbative quantum field theory
as an example how an underlying quantum model can
contribute to the classical one
\cite{birr,barv}. For instance, quantum models of
inflationary cosmology are selected by very
particular properties of observable (classical) Universe
\cite{linde}. It also seems that the main criterion of
selecting quantum gravitation theories is their
appropriate 'dequantization' up to low energy (quantum and
classical) physics \cite{asht}. 

Of course, our exposition is far from covering the whole scope
of {\it International Journal of Geometric Methods in Modern
Physics}. For instance, we are not concerned here with
quantization of geometric objects such as loop quantum gravity
[25-28], strings and branes [29-38]. 
It is also worth mentioning the theory of quantum information
and computation which throws new lights on quantum physics
[39-41].  Geometric techniques in this theory
(e.g., quantum holonomy computation) are intensively developed
\cite{fujii,pach}.

\glos{I.}

Let us start from familiar differential geometry. There
are the following reasons why this geometry contributes to
quantum theory. 

(i) Most of the quantum models comes from
quantization of the original classical
systems and, therefore, inherits their differential geometric
properties. First of all, this is the case of canonical
quantization  which replaces the Poisson bracket
$\{f,f'\}$ of smooth functions with the bracket
$[\wh f,\wh f']$ of Hermitian operators in a Hilbert space such
that Dirac's condition
\be
[\wh f,\wh f']=-i\hbar\wh{\{f,f'\}} 
\ee
holds. Let us mention Berezin--Toeplitz quantization
\cite{engl,rawn,bord} and geometric quantization 
[10,46-48] of
symplectic, Poisson and K\"ahler manifolds. 
This is also the case of quantum gauge theory whose generating
functional is expressed in the classical action of gauge
fields containing topological terms and
anomalies \cite{bert}.  

(ii) Many quantum systems are considered on a smooth manifold
equipped with some background geometry. As a consequence,
quantum operators are often represented by differential
operators which act in a pre-Hilbert space
of smooth functions. A familiar example is the Schr\"odinger
equation. The Kontsevich
deformation quantization is based on the quasi-isomorphism of
the graded differential Lie algebra of multivector fields
(endowed with the Schouten--Nijenhuis bracket and the
zero differential) to that of polydifferential operators
(provided with the Gerstenhaber bracket and the modified
Hochschild differential) \cite{konts,hin}.  
It is also worth reminding perturbative quantum field theory in
curved space.  Firstly, the counter-terms in the energy-momentum
tensor of quantum fields can contribute to the Einstein
equations. This is the case of some models of inflationary
cosmology, e.g.,
its seminal one taking into account the conformal anomaly 
\cite{stra}. Secondly,  different counter-terms in an
effective action can be computed
\cite{birr,barv,pron}. In the case of Riemannian manifolds
(with boundaries), the heat kernel technique is applied
[54-56]. It is based on spectral properties of
Laplace-type and Dirac-type operators on compact manifolds
\cite{gilk95,esp98}.

(iii) In some quantum models, differential geometry is called
into play as a technical tool. For instance, a
suitable $U(1)$-principal connection is used in order to
construct the operators
$\wh f$ in the framework of geometric quantization. Another
example is Fedosov's deformation quantization where a 
symplectic connection plays a similar role \cite{fed}.
Note that this application has stimulated the  study of
symplectic connections \cite{gelf}.

(iv) Geometric constructions in quantum models often
generalize the classical ones, and they are build in a
similar way. For example, connections on principal
superbundles \cite{bart}, graded principal bundles
\cite{stavr}, and quantum principal bundles
\cite{majid2} are defined by means of the
corresponding one-forms in the same manner as connections on
smooth principal bundles with structure finite-dimensional Lie
groups.

\glos{II.}

In quantum models, one deals with infinite-dimensional smooth
Banach and Hilbert manifolds and (locally trivial) Hilbert and
$C^*$-algebra bundles. 
The definition of smooth Banach (and Hilbert) manifolds 
follows that of finite-dimensional smooth manifolds in general,
but infinite-dimensional Banach manifolds are not locally
compact and paracompact \cite{vais73,lang95}. In particular, a
Banach manifold admits the differentiable partition of unity
iff its model space does. 
It is essential that 
Hilbert manifolds (but not,
e.g.,  nuclear (Schwartz) manifolds) satisfy
the inverse function theorem and, therefore, locally trivial
Hilbert bundles are defined. However, they need not be bundles
with a structure group.
 
(i) Infinite-dimensional K\"ahler manifolds provide
an important example of
Hilbert manifolds 
\cite{muj}. In particular, the projective Hilbert
space of complex rays
in a Hilbert space $E$ is such a K\"ahler manifold. This 
is the space the pure states of a
$C^*$-algebra $A$ associated to the same irreducible 
representation $\pi$ of
$A$ in a Hilbert space $E$ \cite{dixm}. Therefore, it
plays
a prominent role in many quantum models. For
instance, it has been 
suggested to consider a loop in the projective
Hilbert space, instead of a parameter space, in order
to describe Berry's phase \cite{anan,bohm}.
We have already mentioned the dequantization procedure which
represents a unital
$C^*$-algebra by a Poisson algebra of
complex smooth functions on a projective Hilbert space
\cite{cir}.

(ii) Gauge theory on a principal bundle $P\to X$ with a
structure compact semisimple matrix Lie group
$G$ over an oriented compact smooth manifold $X$ also provides
an example of Hilbert manifolds. Namely, the
Sobolev $k$-completions
$\gG_k$, $\ol\gG_k$ and $\gG^0_k$  of the gauge group, the 
effective gauge group, and the pointed gauge group,
respectively, are infinite-dimensional Lie groups,  while the
Sobolev completion
$\gC_k$ of the space of principal connections and that $\ol
\gC_k$ of irreducible connections are a Hilbert manifold and its
dense open subset, respectively \cite{mitt}. If
$k>\di X/2+1$, 
the orbit space 
$\ol\gM_k=\ol\gC_k/\ol\gG_{k+1}$ of principal connections is a
smooth Hilbert manifold, and
$\ol\gC_k\to\ol\gM_k$
is a principal bundle 
with the structure Lie group $\ol\gG_{k+1}$. The quotient
$\gC_k/\gG_{k+1}^0$ is also a smooth Hilbert manifold, whereas 
$\gM_k=\gC_k/\gG_{k+1}$ falls
into a countable union of Hilbert manifolds
\cite{kond,vol}. For instance, if the principal bundle
$\ol\gC_k\to\ol\gM_k$ is not
trivial, one meets the well-known Gribov ambiguity \cite{sing}.
It is also interesting that connections on the
$G$-principal bundle
$P\times
\ol\gC\to X\times\ol\gC$ restart the BRST transformations of the
geometric sector of the Donaldson theory \cite{birm} and those
by Witten \cite{witt88}.

(iii) A Hilbert bundle  over a smooth finite-dimensional
manifold $X$ is a particular locally trivial continuous
field of Hilbert spaces in \cite{dixm}. Conversely, one can
think of any locally trivial continuous fields of Hilbert
spaces and $C^*$-algebras as being topological fiber bundles.
Given a Hilbert
space $E$, let $B\subset B(E)$ be some $C^*$-algebra of bounded
operators in $E$. The following fact reflects the
nonequivalence of Schr\"odinger and Heisenberg
quantum pictures. There is the obstruction to the
existence of associated (topological) Hilbert and
$C^*$-algebra bundles $\cE\to X$ and
$\cB\to X$ with the typical fibers $E$ and $B$, respectively.
Firstly,  transition functions of $\cE$ define those
of $\cB$, but the latter need not be continuous, unless 
$B$ is the algebra of compact
operators in $E$.
Secondly, transition functions of $\cB$ need
not give rise to transition functions of $\cE$.
This obstruction is
characterized by the Dixmier--Douady class of $\cB$ in the \v
Cech cohomology group $H^3(X,\Bbb Z)$. There is the similar
obstruction to the $U(1)$-extension of structure groups of 
principal bundles \cite{bryl,car}. One also meets the
Dixmier--Douady class as the obstruction to a bundle gerbe
being trivial \cite{bouw}.

(iv) There is a problem in the definition of a connection on
$C^*$-algebra bundles which comes from the
fact that a $C^*$-algebra (e.g., any commutative
$C^*$-algebra) need not admit nonzero bounded derivations. An
unbounded derivation  of a $C^*$-algebra $A$ obeying certain
conditions is an infinitesimal generator of a strongly (but not
uniformly) continuous one-parameter group of automorphisms of
$A$ \cite{brat75}. Therefore, one may introduce a connection 
on a $C^*$-algebra bundle in terms of
parallel transport curves and operators, but not their
infinitesimal generators \cite{asor}. Moreover, a representation
of $A$ does not imply necessarily a unitary representation of
its strongly (not uniformly) continuous one-parameter group of
automorphisms. In contrast, connections on a Hilbert bundle over
a smooth manifold can be defined both as particular first order
differential operators on the module of its sections
\cite{book00} and a parallel displacement along paths
lifted from the base \cite{iliev}. 

(v) Instantwise geometric
quantization of time-dependent mechanics is phrased in
terms of Hilbert bundles over $\Bbb R$ \cite{sni,jmp02a}.
Holonomy operators in a Hilbert bundle with a
structure finite-dimensional Lie group are well known to
describe the non-Abelian geometric phase phenomena
\cite{bohm03}. At present, holonomy operators in Hilbert
bundles attract special attention in connection with quantum
computation and control theory \cite{fujii,pach,jmp04}.

\glos{III.}

Geometry in
quantum systems speaks mainly the algebraic language of rings, 
modules and sheaves due to the fact that the basic
ingredients in the differential calculus and differential
geometry on smooth manifolds (except nonlinear differential
operators) can be restarted in a pure algebraic way. 

(i) Any smooth real manifold $X$ is homeomorphic to
the real spectrum of the $\Bbb R$-ring $C^\infty(X)$ of
smooth real functions on $X$ provided with the Gelfand topology
\cite{atiy,nestr}. Furthermore, the sheaf
$C^\infty_X$ of germs of $f\in C^\infty(X)$ on
this topological space fixes a unique smooth
manifold structure on $X$ such that it is the sheaf
of smooth functions on $X$. The pair $(X,C^\infty_X)$
exemplifies a local-ringed space. A sheaf 
$\gR$ on a topological space $X$ is said to be a 
local-ringed space if its
stalk $\gR_x$ at each point $x\in X$ is a local commutative 
ring \cite{tenn}. One can associate to any
commutative ring $\cA$ the particular local-ringed space,
called an affine scheme, on the spectrum Spec$\,\cA$ of $\cA$
endowed with the Zariski topology \cite{ueno}. 
Furthermore, one can assign the following
algebraic variety to any commutative finitely
generated $\cK$-ring
$\cA$ over an algebraically closed field $\cK$. Given a ring 
$\cK[x]$ of polynomials with
coefficients in
$\cK$, let us
consider the epimorphism $\phi: \cK[x]\to\cA$ defined by 
the equalities
$\phi(x_i)=a_i$, where 
$a_i$ are generating  elements of $\cA$.  Zeros of
polynomials in Ker$\,\phi$ make up an algebraic variety $\cV$
whose coordinate ring $\cK_\cV$ is exactly
$\cA$. The subvarieties of $\cV$ constitute the system of
closed sets of the Zariski topology on $\cV$ \cite{shaf}.
Every affine variety $\cV$ in turn yields the
affine scheme Spec$\cK_\cV$ such that there is one-to-one
correspondence between the points of Spec$\cK_\cV$ and the
irreducible subvarieties of $\cV$. For instance, complex
algebraic varieties have a structure of complex analytic
manifolds.

(ii) Given a (connected) compact topological space $X$ and 
the ring $\Bbb C^0(X)$ of continuous complex functions on $X$,
the well-known Serre--Swan theorem
\cite{swan} states  that a
$\Bbb C^0(X)$-module is finitely generated projective iff it
is isomorphic to the module of sections of some (topological)
vector bundle over $X$. Moreover, this isomorphism is a
categorial equivalence \cite{karo}, and its variant takes place
if $X$ is locally compact 
\cite{ren}. If $X$ is a compact smooth manifold, there is the
similar isomorphism of a finitely generated
projective
$C^\infty(X)$-modules on  $X$ to the
modules of sections of some smooth vector bundle over $X$
\cite{var}, and this is also true if
$X$ is not necessarily compact. A variant of the Serre--Swan
theorem for Hilbert modules over noncommutative $C^*$-algebras
holds \cite{kawa}.

(iii) Let $\cK$ be a commutative ring, $\cA$ a
commutative 
$\cK$-ring, and $P$, $Q$ some $\cA$-modules. The
$\cK$-linear $Q$-valued differential operators on
$P$ can be defined
\cite{nestr,grot,vinb}. The representative objects of the
functors
$Q\to \dif_s(P,Q)$ are the jet modules $\cJ^sP$ of $P$. 
Using the first order jet module $\cJ^1P$, one also restarts the
notion of a connection on a $\cA$-module $P$
\cite{book00,kosz60}. Such a connection assigns to each
derivation
$\tau\in\gd\cA$ of a $\cK$-ring $\cA$ a first order
$P$-valued differential operator $\nabla_\tau$ on $P$ 
obeying the Leibniz rule
\be
\nabla_\tau(ap)=\tau(a)p+a\nabla_\tau(p).
\ee 
For instance, if $P$ is a $C^\infty(X)$-module of
sections of a smooth vector bundle $Y\to X$, we come to the
familiar notions of a linear differential operator on $Y$,
the jets of sections of $Y\to X$ and a linear connection on
$Y\to X$. Similarly, connections on 
local-ringed spaces are introduced \cite{book00}. In
supergeometry, connections on graded modules  over a
graded commutative ring and graded local-ringed spaces are
defined \cite{bart}. In noncommutative geometry,
different definitions of a differential operator on
modules over a noncommutative ring have been suggested
[99-101]. Roughly speaking, 
the difficulty lies in the fact that, if $\dr$
is a derivation of a noncommutative $\cK$-ring $\cA$, the
product $a\dr$, $a\in\cA$, need not be so. There are
also different definitions of a connection on modules over a
noncommutative ring \cite{dub,land}.

(iv) Let $\cK$ be a commutative ring, $\cA$ a
(commutative or noncommutative)
$\cK$-ring, and $\cZ(\cA)$ the center of $\cA$. Derivations of
$\cA$ make up a Lie
$\cK$-algebra
$\gd\cA$. Let us consider the Chevalley--Eilenberg complex of
$\cK$-multilinear morphisms of $\gd\cA$ to $\cA$, seen as a
$\gd\cA$-module \cite{vais,fuks}. Its subcomplex
$\cO^*(\gd\cA,d)$ of
$\cZ(\cA)$-multilinear morphisms is a graded differential
algebra, called the Chevalley--Eilenberg differential calculus
over
$\cA$. It contains the universal differential calculus
$\cO^*\cA$ generated by elements $da$, $a\in\cA$. If $\cA$ is
the $\Bbb R$-ring $C^\infty(X)$ of smooth real functions on a
smooth manifold $X$, the module $\gd C^\infty(X)$ of its
derivations is the Lie algebra of vector fields on $X$ and
the Chevalley--Eilenberg differential calculus over
$C^\infty(X)$ is exactly the algebra of
exterior forms on a manifold $X$ where the Chevalley--Eilenberg
coboundary operator $d$ coincides with the exterior
differential, i.e.,  
$\cO^*(\gd C^\infty(X),d)$ is the familiar de Rham complex.
In a general setting, one therefore can
think of elements of the Chevalley--Eilenberg differential
calculus $\cO^k(\gd\cA,d)$ over an algebra $\cA$ as being
differential forms over
$\cA$. Similarly, the Chevalley--Eilenberg differential
calculus over a graded commutative ring is constructed
\cite{fuks}. 

\glos{IV.}

As was mentioned above, homology and cohomology of spaces and
algebraic structures often play a role of {\it
sui generis} hidden quantization parameters which can
characterize nonequivalent quantizations. 

(i) First of all, let us mention the abstract de Rham theorem 
\cite{hir} and, as its corollary, the homomorphism $H^*(X,\Bbb
Z)\to H^*(X)$ of the \v Cech cohomology of a smooth manifold
$X$ to the de Rham cohomology of exterior forms on $X$. 
For instance, the Chern classes $c_i\in H^{2i}(X,\Bbb
Z)$ of a $U(n)$-principal bundle $P\to X$ are represented by the
de Rham cohomology classes of certain characteristic exterior
forms $\cP_{2i}(F_A)$ on
$X$ expressed in the strength two-form $F_A$ of a principal
connection $A$ on $P\to X$ \cite{eguchi}.  The Chern class $c_2$
of a complex line bundle plays a prominent role in many
quantization  schemes, e.g., geometric quantization. 
Conversely, given a principal bundle $P\to X$ with a structure
Lie group
$G$, the Weil homomorphism  associates to any invariant
polynomial
$I_k$ on the Lie algebra of
$G$ a closed exterior form $\cP_{2k}(F_A)$ on $X$ whose
de Rham cohomology class is independent of the
choice of a connection $A$ on $P$. Furthermore, 
given another principal connection
$A'$ on $P$, the  global transgression formula
\be
\cP_{2k}(F_A)-\cP_{2k}(F_{A'})=dS_{2k-1}(A,A')
\ee
defines the secondary characteristic form $S_{2k-1}(A,A')$. In
particular, if $\cP_{2k}$ is the characteristic Chern form, then
$S_{2k-1}(A,A')$ is the familiar Chern--Simons $(2k-1)$-form
utilized in many models of topological field theory
\cite{birm,hu} and (gauge and BRST) anomalies \cite{bert}.

(ii) The
well-known index theorem establishes the equality of the
index of an elliptic operator on a fiber bundle to its
topological index expressed in terms of the characteristic
forms of the Chern character,  Todd and Euler classes. 
Note that the classical
index theorem deals with linear elliptic operators on
compact manifolds. They are
Fredholm  operators. In order to 
generalize the index theorem to noncompact manifolds, 
one either imposes
conditions sufficient to force operators to be the
Fredholm ones or considers the operators which are no longer
Fredholm, but their index can be interpreted as a real number 
by some kind of averaging procedure \cite{roe}.
The index problem 
for nonlinear elliptic operators has also been discussed
\cite{palais}. From the physical viewpoint, the index theorem 
has shown that topological invariants can be expressed in
terms of functional analysis, e.g., functional integrals in
field models \cite{birm,witt}. The following model is
quite illustrative. Given two circles $s, s'\in\Bbb R^3$, their
linking number obeys the well-known integral formula 
\be
L(s,s')=(4\pi)^{-1}\op\int_s dx^i\op\int_{s'}dy^j \ve_{ijk}
\dr_k[\op\sum_r(x^r-y^r)^2]^{-1/2}.
\ee
This formula is generalized to compact homologically trivial
surfaces of dimensions
$p$ and $n-p-1$ in $\Bbb R^n$. Let us now consider the
topological field theory of a $p$-form $B$ and a
$(n-p-1)$-form $A$ on a compact oriented $n$-dimensional
manifold $Z$ (without boundary). Its action is 
\be
S(A,B)=\op\int_Z B\w dA.
\ee 
It is invariant under adding an
exact form either to $A$ or $B$. Then the linking number of compact homologically
trivial surfaces $M$ and $N$  of dimensions
$p$ and $n-p-1$, respectively, can be given by the functional
integral
\be
L(M,N)=i^{-1}\int\m(A,B)\exp(iS(A,B))
\ee
by the appropriate choice of the measure $\m(A,B)$ on the
moduli space of fields $A$ and $B$ \cite{horow}. This is the
simplest example of abelian and non-Abelian topological $BF$
theories of antisymmetric tensor fields \cite{szabo}. For
instance, polynomial invariants of knots are  obtained in
three-dimensional versions of $BF$ theory
\cite{catt}, where a knot $K=\dr\Si_K$ is represented by
a (classical) observable
\be
\g^{(\Si_K, K, x_0)}=\op\int_{\Si_K} B\w A
+\frac12\op\oint_{x<y\in K} [A(x)B(y)-B(x)A(y)].
\ee
A non-Abelian $BF$ theory also serves as a dual
model for Yang--Mills theory \cite{catt2}. 
 
 (iii) Several important characteristics come from geometry and
topology of moduli spaces. They are exemplified by  Donaldson
and  Seiberg--Witten invariants. 
Given an $SU(2)$-principal
bundle $P\to X$ over a compact four-dimensional manifold $X$, we
have the
$SU(2)$-principal bundle $(P\times \ol\gC)/\ol\gG\to
X\times\ol\gM$,
where the appropriate Sobolev completion is assumed. By the
K\"unneth formula its second Chern class $c_2\in
H^4(X\times\ol\gM)$ is decomposed into the terms  
$c_{i,4-i}\in H^i(X)\ot H^{4-i}(\ol\gM)$, $i=0,\ldots,4$,
which provide the map $\op\times^m H_2(X;\Bbb Z)\to
H^{2m}(\ol\gM,\Bbb Q)$
via the cup product in $H^{2m}(\ol\gM,\Bbb Q)$. This map
yields an injection
of the polynomial algebra on $H_2(X;\Bbb Z)$ to $H^{\rm
even}(\ol\gM,\Bbb Q)$.
Evaluated on the homology cycle
$[\cM]\in H_*(\ol\gM,\Bbb Q)$ of the module space $\cM$ of
irreducible instantones, these polynomials are
the Donaldson polynomials \cite{don}. Expressed in
the strength of a connection on the irreducible orbit space
$\ol\gM$, they are the differential Donaldson invariants
 of a four-dimensional manifold $X$.
Similarly, let $X$ be an oriented compact four-dimensional
Riemannian manifold provided with a spin$^c$ structure, $S\to
X$ the corresponding positive spinor bundle, and $L\to X$ the
associated complex line bundle. Let $\gS$ be an appropriate
Sobolev completion of the irreducible configuration space of
connections on $L\to X$ and sections of
$S\to X$, except the zero section.
Given the Sobolev completions $\gG$ and $\gG^0$ of the
$U(1)$-gauge group and its pointed subgroup, respectively, we
have the
$U(1)$-principal bundle $\gS/\gG^0\to \gS/\gG$. The
Seiberg--Witten invariant is defined as the cup-product
$\op\smile^d c_1$ of the first Chern class of this bindle
evaluated on the homology class the module space $\cM_\f$ of
solutions of the perturbed Seiberg--Witten equations ($d=\di
\cM_\f/2$) \cite{moor}.  

(iv) Geometric quantization of a symplectic manifold
$(X,\Om)$ is affected by the following ambiguity.
Firstly, the equivalence classes of admissible connections on a
prequantization bundle (whose curvature obeys the
prequantization condition $R=i\Om$) are indexed by the set
of homomorphisms of the homotopy group $\pi_1(X)$ of $X$
to
$U(1)$  \cite{kost,mck}. Secondly, there are nonequivalent
bundles of half-forms over $X$ in general and, consequently, the
nonequivalent quantization bundles \cite{eche98}. This
ambiguity leads to nonequivalent quantizations.

(v) The cohomology analysis gives a rather complete picture of
deformation quantization of symplectic manifolds. Let $\cK$ be
a commutative ring and $\cK[[\la]]$ the ring of formal series
in a real parameter $\la$. Recall that, given an associative
(resp. Lie) algebra
$A$  over a commutative ring $\cK$, its Gerstenhaber
deformation \cite{gest} is an associative (resp. Lie) 
$\cK[[\la]]$-algebra $\ol A$ such that $\ol A/\la\ol A\ap A$.
The multiplication in $\ol A$ reads
\be
a\star b= a\circ b+\op\sum_{r=1}^\infty \la^rC_r(a,b)
\ee
where $\circ$ is the original associative (resp. Lie) product 
and $C_r$ are 2-cochains of the Hochschild (resp.
Chevalley--Eilenberg) complex of $A$. The obstruction to the
existence of a deformation of $A$ lies in the
third Hochschild (resp.
Chevalley--Eilenberg) cohomology group. Let now $A=\Bbb
C^\infty(X)$ be the ring of complex smooth functions on a
smooth manifold $X$. One  considers its associative
deformations $\ol A$ where the cochains $C_r$ are
bidifferential operators of finite order. The multidifferential
cochains make up a subcomplex of the Hochschild complex of $A$,
and its cohomology equals the space of multi-vector fields on
$X$ \cite{vey}. If $\Bbb C^\infty(X)$ is provided with the
standard Fr\'echet topology of compact convergence for 
all derivatives, one can consider its
continuos deformation. The corresponding subcomplex of the
Hochschild complex of $A$ is proved to have the same cohomology
as the differential one
\cite{nad}. Let now $X$ be a
symplectic manifold, and let $A=\Bbb C^\infty(X)$ be the Poisson
algebra. Since the Poisson bracket is a bidifferential
operator of order (1,1), one has studied the similar
deformations of $A$ where the cochains $C_r$ are differential
operators of order (1,1) with no constant term. The cohomology
of the corresponding subcomplex of the
Chevalley--Eilenberg complex of $A$ equals the de
Rham cohomology
$H^*(X)$ of
$X$ \cite{lich}. 
The equivalence classes of Poisson deformations of the Poisson
bracket on a symplectic manifold $X$ are parametrized by
$H^2(X)[[\la]]$. A
star-product on a Poisson manifold is defined as an
associative deformation of $\Bbb C^\infty(X)$ such that 
$C_1(f,f')-C_1(f',f)$ is the Poisson bracket. The existence of
a star product on an arbitrary symplectic manifold
has been proved in \cite{wild}, and this is true
for any regular Poisson manifold \cite{masm,fed1}.
Moreover, any star-product on a symplectic manifold is
equivalent to Fedosov's one, and its equivalence classes are
parametrized by  $H^2(X)[[\la]]$ \cite{nest,gutt}.

(vi) Let us also mention BRST cohomology, called into play
in order to describe constrained symplectic systems
[125-127]. Let $(Z,\Om)$ be a symplectic
manifold endowed with a Hamiltonian action of a Lie group
$G$, $\wh J$ the corresponding momentum mapping of $Z$ to the
Lie coalgebra $\cG^*$ of $G$, and $N=\wh J^{-1}(0)$ a regular
constraint surface. The classical BRST complex is defined as
the bicomplex 
\be
B^{n,m}=\op\w^n\cG^*\ot\op\w^m\cG\ot
C^\infty(Z), 
\ee
where the $n$- and $m$-gradings are the
ghost and antighost degrees, respectively. The 
differential $\dl: B^{*,*}\to B^{*+1,*}$ is the coboundary
operator of the Chevalley--Eilenberg cohomology of $\cG$ with
coefficients in the $\cG$-module $\cG\ot C^\infty(Z)$, while 
$\dr: B^{*,*}\to B^{*,*-1}$ is the Koszul boundary operator. The
total differential $d=\dl+2\dr$ is the classical BRST
operator. The algebra $B$ is provided with the super Poisson
bracket $\{,\}$, and there exists an element $\Theta$ of $B$,
called the BRST charge, such that $d=\{\Theta,.\}$. The
BRST cohomology is defined as the cohomology of $d$. The BRST
complex has been built for constrained
Poisson systems \cite{kimura} and time-dependent Hamiltonian
systems with Lagrangian constraints \cite{mang00} as an
extension of the Koszul--Tate complex of constraints through
introduction of ghosts. Quantum BRST cohomology has been studied
in the framework of geometric \cite{tuyn} and
deformation \cite{bord1} quantization. In field theory, BRST
transformations are the particular generalized supersymmetry
transformations of first order (depending on jets of
ghosts) whose infinitesimal generator (the BRST operator) is
nilpotent. In the framework of the Lagrangian field-antifield
formalism on a base space $X$
\cite{barn,bran01}, one considers the bicomplex of horizontal
(local in the terminology of \cite{barn}) exterior forms on the
infinite order jet space of fields, ghosts and anti-fields with
respect to the BRST operator ${\bf s}$ and the horizontal
differential $d_H$. In particular, it is essential for
applications that relative (${\bf s}$ modulo $d_H$) cohomology
of this bicomplex in the maximal form-degree $n=\di X$ are
related to the cohomology of the total BRST operator ${\bf
s}+d_H$ via the descent equation. Stated for a contractible
base space $X$, this relation can be extended to an arbitrary
manifold $X$ due to the fact that the cohomology of
$d_H$ equals the de Rham cohomology of $X$ \cite{lmp}.

\glos{V.}

Contemporary quantum models appeal to a number of new algebraic
structures and the associated geometric techniques.

(i) For instance, SUSY models deal with graded
manifolds and different types of supermanifolds, namely, 
$H^\infty$-, $G^\infty$-, $GH^\infty$-, $G$-supermanifolds
over (finite) Grassmann algebras, $R^\infty$- and
$R$-supermanifolds over Arens--Michael algebras of Grassmann
origin and the corresponding types of DeWitt supermanifolds  
\cite{bart,bart93,bruz99}. Their geometries are phrased in
terms of graded local-ringed spaces. Note that one usually
considers supervector bundles over
$G$-supermanifolds.  Firstly, the category of these supervector
bundles is  equivalent to the
category of locally free sheaves of finite rank (in contrast,
e.g., with $GH^\infty$-supermanifolds). Secondly, derivations
of the structure sheaf of a
$G$-supermanifold  constitute a locally free sheaf (this is
not the case, e.g., of $G^\infty$-supermanifolds).
Moreover,  this sheaf is again a structure sheaf of
some $G$-superbundle (in contrast with graded manifolds). 
At the same time, most of the quantum models uses graded
manifolds. They are not supermanifolds, though there is the
correspondence between graded manifolds and DeWitt
$H^\infty$-supermanifolds. By virtue of the well-known 
Batchelor theorem, the structure ring of any graded
manifold with a body manifold $Z$ is isomorphic to
the graded ring $\cA_E$ of sections of some exterior bundle $\w
E\to Z$. In physical models, this isomorphism holds fixed
from the beginning as a rule and, in fact, by geometry of a
graded manifold is meant the geometry of the graded ring
$\cA_E$. For instance, the familiar differential calculus
in graded exterior forms is the graded
Chevalley--Eilenberg differential calculus over such a ring.

(ii) Noncommutative geometry is mainly developed
as a generalization of the calculus in 
commutative rings of
smooth functions \cite{conn,land,grac}. 
In a general setting, any
noncommutative $\cK$-ring
$\cA$ over a commutative ring $\cK$ can be called into play. One
can consider the above mentioned  Chevalley--Eilenberg
differential calculus
$\cO^*\cA$ over
$\cA$, differential operators and connections on 
$\cA$-modules (but not their jets). If the derivation
$\cK$-module
$\gd\cA$ is a finite projective module with respect to the
center of
$\cA$, one can treat the triple $(\cA,\gd\cA,\cO^*\cA)$ as a
noncommutative space. For instance, this is the case of the
matrix geometry, where $\cA$ is the algebra of finite matrices, 
and of the quantum phase space,  where $\cA$ is a
finite-dimensional CCR algebra. Noncommutative
field theory also can be treated in this manner
\cite{doug,poly}, though the bracket of space coordinates
$[x^\m,x^\nu]=i\th^{\m\nu}$ in this theory is also restarted
from Moyal's star product $x^\m\star x^\nu$
\cite{doug,shiwu}. A different linear coordinate product
$[x^\m,x^\nu]=ic^{\m\n}_\la x^\la$ comes from Connes'
noncommutative geometry \cite{marmo}. In Connes'
noncommutative geometry, the more deep analogy to the
case of commutative smooth function rings leads to the notion
of a spectral triple
$(\cA,E,\cD)$ 
\cite{conn,conjmp}. It is given by an involutive subalgebra
$\cA\subset B(E)$ of bounded operators on a Hilbert space
$E$ and an (unbounded) self-adjoint operator $\cD$ 
in
$E$ such that the resolvent $(\cD-\la)^{-1}$, $\la\in\Bbb
C\setminus\Bbb R$ is a compact operator and
$[\cD,\cA]\subset B(E)$.  Furthermore, one 
assigns to elements $\f=a_0da_1\cdots da_k$ of the universal
differential calculus $(\cO^*\cA,d)$ over $\cA$ the operators 
$\pi(\f)=a_0[\cD,a_1]\cdots [\cD,a_k]$ in $E$. This however
fails to be a representation of the differential
algebra $\cO^*\cA$ because
$\pi(\f)=0$ does not imply $\pi(d\f)=0$. The appropriate
quotient
$\cO^*\cA\ni\f\to[\f]\in\cO^*_\cD$ together with the
differential $\dl[\f]=[d\f]$ overcomes this difficulty, though
$\pi([\f])$ is not an operator in $E$. 
The 
$(\cO^*_\cD,\dl)$ is called  
the Connes--de Rham complex. Note that other variants of
spectral data, besides a spectral triple, are also discussed
\cite{frol1}. The algebra $C^\infty(X)$ and the
Dirac operator $\cD$ on a compact manifold $X$ exemplifies
Connes' commutative geometry
\cite{conn,ren2}.  Spectral triples have
been studied for noncommutative tori, the Moyal deformations of
$\Bbb R^n$,  noncommutative spheres $2$-,$3$- and $4$-spheres
\cite{con02,chak}, and quantum Heisenberg manifolds
\cite{chak2}. It is essential that,  since
the universal differential calculus over a unital commutative
algebra $\cA$ 
is a direct summand (not as a differential algebra) of
the Hochschild homology of $\cA$
\cite{loday}, the differential calculus $\cO^*\cA$
over $\cA$ in Connes'
noncommutative geometry is generated by the Hochschild cycles
of $\cA$ so that the representation
$\pi$ of Hochschild boundaries $b$ vanishes.

(iii) Formalism of groupoids provides the above mentioned
categorial $C^*$-algebraic deformation quantization of some
class of Poisson manifolds \cite{lands02,lands01}. A
groupoid is a small category whose morphisms are
invertible \cite{rena,mack}. For instance, given  an action of
a group $G$ on a set $X$ on the right, the product
$\gG=X\times G$ is brought into the action groupoid, where 
a pair $((x,g),(x',g'))$ is composable iff $x'=xg$, 
the inversion $(x,g)^{-1}:=(xg,g^{-1})$,
the partial multiplication $(x,g)(xg,g'):=(x,gg')$,
the range $r((x,g)):=(x,1_G)$, and 
the domain $l((x,g)):=(xg,1_G)$.
The unit space $\gG^0=r(\gG)=l(\gG)$ of this groupoid is naturally
identified to $X$. Any
group
bundle $Y\to X$ (e.g., a vector bundle) is a groupoid
whose elements make up  composable pairs iff they belong to the
same fiber, and whose  unit space is the set of unit elements of
fibers of $Y\to X$. Let $\gA\to\gG^0$ be an Abelian 
group bundle over the unit space $\gG^0$ of a
groupoid $\gG$. The pair
$(\gG,\gA)$ together with a homomorphism $\gG\to {\rm
Iso}\,\gA$ is called the 
$\gG$-module bundle. One can associate to any $\gG$-module
bundle a cochain complex $C^*(\gG,\gA)$. Let $\gA$ be a
$\gG$-module bundle in groups $U(1)$. The key point is
that, similarly to the
case of a locally compact group \cite{dixm},  one can
associate a
$C^*$-algebra $A_{\gG,\si}$ to any locally compact groupoid
$\gG$ provided with a  Haar system by means of the choice of a
two-cocycle
$\si\in C^2(\gG,\gA)$ \cite{rena}. The algebras $A_{\gG,\si}$
and $A_{\gG,\si'}$ are isomorphic if $\si$ and $\si'$ 
are cohomology equivalent. If $\gG$ is an $r$-discrete groupoid,
any measure $\la$ of total mass 1 on its unit space $\gG^0$
induces a state of the $C^*$-algebra
$A_{\gG,\si}$. Moreover, it is the KMS state if a measure $\la$
satisfies a certain cohomological condition. A Lie groupoid is 
a groupoid for which $\gG$ and
$\gG^0$ are smooth manifolds, the inversion and partial
multiplication are smooth, while $r$ and $l$
are fibered manifolds. Since a Lie groupoid admits a Haar
system, one can assign to it a $C^*$-algebra
$A_\gG$. This assignment is functorial 
if certain classes of morphisms of Lie groupoids and
$C^*$-algebras (isomorphism classes of regular bibundles
and those of Hilbert bimodules, respectively)
are considered \cite{lands01}. A Lie groupoid  is called
symplectic if it is a symplectic manifold
$(\gG,\Om)$ such that the multiplication relation $(x,y)\to
(xy,x,y)$  is a Lagrangian submanifold of the
symplectic manifold
$(\gG\times\gG\times\gG,\Om\ominus\Om\ominus\Om)$ \cite{bates}.
A Poisson manifold $P$ is called integrable if there exists a
symplectic groupoid $\gG(P)$ over $P$. It
is unique up to an isomorphism. Integrable Poisson manifolds
subject to a certain class morphisms (isomorphism classes of
regular dual pairs) make up a  suitable category {\bf Ps}
\cite{lands01}. Since the groupoid
$\gG(P)$ is
$l$- and $l$-simple connected, one considers the
category 
{\bf LGc} of Lie groupoids  possessing this property. Any
Lie groupoid yields an associated Lie algebroid $L(\gG)$ which
is the restriction to $\gG^0$ of the vertical tangent bundle
of the fibration
$r:\gG\to\gG^0$   \cite{mack}. Sections of $L(\gG)$
make up a real Lie algebra compatible with the anchor map
$L(\gG)\to T\gG^0$. The key point is that, similarly to the
dual of a Lie algebra, the dual
$L^*(\gG)$ of $L(\gG)$ is a Poisson manifold in a canonical
way. Then the assignment $\gG\mapsto L^*(\gG)$ is a functor
 ${\bf LGc} \to {\bf Ps}$ \cite{lands01}. Let
{\bf LPs} denote its image. One can show
that $\cQ: L^*(\gG)\mapsto A_\gG$ is a functor from
the category {\bf LPs} to the above mentioned category of
$C^*$-algebras \cite{lands02}.
It is a desired functorial quantization. This
functor is equivariant under the Morita equivalence  of Poisson
manifolds in {\bf LGc} \cite{xu} and that of $C^*$-algebras 
\cite{rief74}. This property fails if the functor $\cQ$ is
extended to the category {\bf Ps}. In this case, the natural
codomain of $\cQ$ is the category {\bf KK} whose objects are
separable
$C^*$-algebras and morphisms are Kasparov's KK-groups
\cite{black}. Furthermore, the functorial quantization  
$L^*(\gG)\mapsto A_\gG$ is amplified into the above
mentioned strict quantization of $A_\gG$ by an appropriate
continuous field of
$C^*$-algebras over $\Bbb R$ \cite{lands02}.
Connes' tangent groupoid provides an example of such strict
quantization
\cite{lands282,cari99}.

(iv) Hopf algebras and, in particular, quantum 
groups make a contribution to many quantum theories
\cite{majid2,klim,majid}. At the same time, the
development of differential  calculus  and 
differential geometry over these algebras has met difficulties.
Given a (complex or real) Hopf algebra
$H=(H,m,\Delta,\ve, S)$, one introduces the first
order differential calculus (henceforth FODC)
$(\cO^1,d)$ over $H$ as for a
noncommutative algebra. It
is said to be left-covariant if $\cO^1$ possesses
the structure of a left $H$-comodule $\Delta_l:\cO^1\to
H\ot\cO^1$ such that $\Delta_i(adb)=\Delta(a)(\id\ot
d)\Delta(b)$, $a,b\in H$ \cite{klim}. By virtue of
Woronowicz's theorem \cite{woron}, left-covariant 
FODCs are classified by right ideals 
\be
\cR=\{x\in\Ker \ve\,:\, S(x_1)dx_2=0)
\ee
of $H$ contained in the kernel
of its counit. The linear subspace 
\be
T=\{t\in H^*\,:\, t(1)=0,\,\, t(\cR)=0\}
\ee
of the dual $H^*$ is the quantum (enveloping) Lie algebra
(quantum tangent space \cite{heck})
associated to the left-covariant FODC
$(\cO^1,d)$ (see \cite{lunts} for a
general construction of the enveloping algebra
for a noncommutative FODC). A problem lies 
 in the definition of vector fields 
as a sum $a^iu_i$ of invariant vector fields
$u_i(a)=a_1t_i(a_2)$ 
 \cite{ash} because they satisfy the
deformed Leibniz rule deduced from the formula
\be
t_i(ab)=t_i(a)\ve(b) +\op\sum_if_{ij}(a)t_j(b),
\ee
where $(t_i)$ is a basis for $T$ and
$f_{ij}$ are functionals on $H$. One can model the vector fields
obeying such a Leibniz rule by the so called Cartan pairs
\cite{bor97}. These are elements $u$ of the right $H$-dual
$\cO_R$ of
$\cO^1$ together with the morphisms $\wh u:H\ni
a\mapsto =u(da)\in H$ which obey the relations
\be
\wh{(bu)}(a)=b\wh u(a), \qquad \wh u(ba)=\wh u(b)a
+\wh{(ub)}(a).
\ee
Note that the standard 
FODC 
given by the kernel of the product
$\cA\ot\cA\to\cA$ in a noncommutative algebra $\cA$ is not
applied to a commutative algebra because it does not provide the
equality
$adb=(db)a$,
$a,b\in\cA$.  The similar commutative FODC
deals with a certain quotient $\cA\ot\cA/\m^2$ \cite{vinb}. 
Another problem of geometry of Hopf algebras is the
notion of a quantum principal bundle
\cite{majid2,brz,calow}. In the case of Lie groups,
there are two equivalent definitions of a smooth principal
bundle, which is both a set of 
trivial bundles  glued together by means of transition
functions and a bundle provided with the canonical action
of a structure group on the right. In the case of quantum
groups, these two notions of a principal bundle
are not matched, unless the base is a smooth manifold
\cite{pfla,durd}. The first
definition of a quantum principal bundle repeats the
classical one and makes use of the notion of a trivial quantum
bundle, a covering of a quantum space (e.g., by a family of
nonintersecting closed ideals), and its reconstruction from
local pieces \cite{budz} which, however, is not always possible
\cite{calow2}. The second definition of a quantum principal
bundle is algebraic \cite{majid2,brz98}. Let $H$ be a
bialgebra and $P$ a right $H$-comodule algebra with respect
to the coaction $\bt: P\to P\ot H$. Let $M=\{p\in P\,:\,
\bt(p)=p\ot 1_H\}$ be its invariant subalgebra. The triple
$(P,H,\bt)$ is called a quantum principal bundle if the map
\be
{\rm ver}: P\ot_M P\ni (p\ot_M q)\to p\bt(q)\in P\ot H
\ee
is a linear isomorphisms. This condition, called the
Hopf--Galois condition, is a key point of this algebraic
definition of a quantum principal bundle. By some reasons, one
can think of it as being a {\it sui generis} local
trivialization. In particular, let $\pi:P\to H$ be a
Hopf-algebra surjection and
$\bt:=(\id\ot\pi)\Delta$. If the product map $m:\Ker\ve|_M\ot
P\to\Ker\pi$ is a surjection, then $(P,H,\bt)$ is a quantum
principal bundle \cite{majid2}. Many examples of a quantum
principal bundle come from this fact.

(v) Finally, one of the main point of Tamarkin's proof of the
formality theorem in deformation quantization is that,
for any algebra $\cA$ over a field characteristic zero, its
cohomological Hochschild complex and its Hochschild cohomology
are algebras for the same operad. This observation has been the
starting point of 'operad renaissance' \cite{konts1,stash}. 
Monoidal categories provide numerous examples of algebras for
operads. Furthermore, homotopy monoidal categories lead to the
notion of a homotopy monoidal algebra for an operad. In a
general setting, one considers homotopy algebras and weakened
algebraic structures where, e.g., a product operation is
associative up to homotopy \cite{lein}. Their well-known
examples are 
$A_\infty$-spaces and
$A_\infty$-algebras \cite{stash63}. At the same time, the
formality theorem is also applied to quantization of
several algebro-geometric structures such as algebraic
varieties \cite{konts,yek}.

\end{document}